An Audit Framework for
# Adopting AI-Nudging on Children

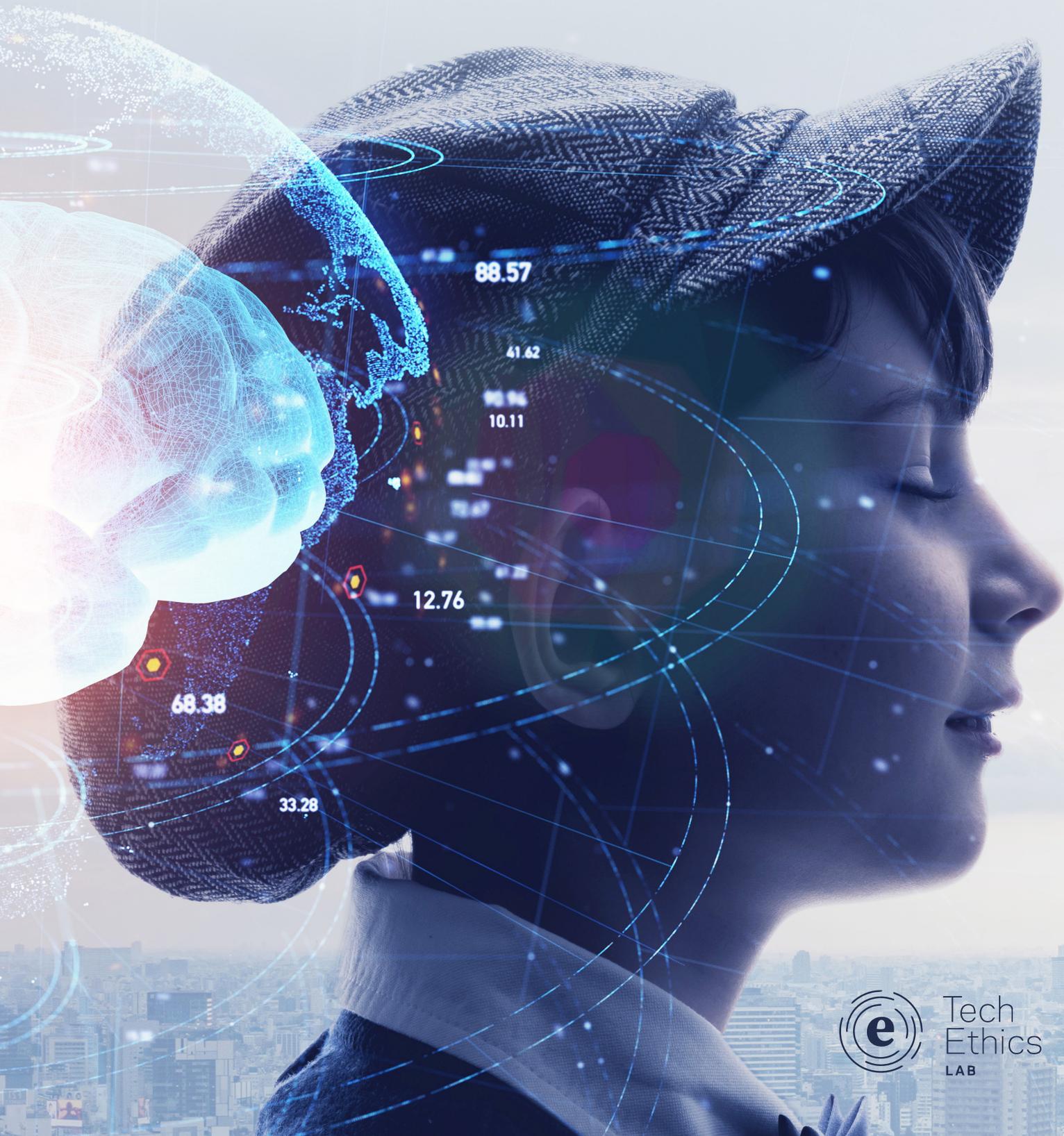

Tech Ethics LAB

# Table of Content

# 1. Introduction

The term "nudges" is used to describe strategic tools designed to influence people's choices and behaviors, usually without subjects consciously recognizing that influence. The idea of "nudge theory" was introduced by Thaler and Sunstein in their 2008 book, Nudge, and is widely studied in both behavioral economics and psychology.

> At times, technological nudges are called "behavioral design" or "algorithmic nudging".

To date, work on nudge theory in technology has focused mostly on choice architectures, designed by user experience (UX) experts, which have mostly predictable outcomes. However, artificial intelligence (AI) has revolutionized the world of nudging: today, an individual's decisions can be shaped by AI interfaces that constantly adapt and change according to the user's choices and detectable behavior. The outcomes in this process are myriad and unpredictable, and nudging may even be an unintended consequence, distinct from the original intended design for the AI interfaces.

> Companies are increasingly using algorithms to manage and control individuals not by force, but rather by nudging them into desirable behavior - in other words, learning from their personalized data and altering their choices in some subtle way. (Möhlmann, 2021)

A detailed discussion of this form of AI-enhanced nudges is currently lacking. At present, commercial legislation (EU Unfair Commercial Practices Directive in the European Union, 2005) and design guidelines (La Forme Des Choix, 2019) exist, which are helpful in mitigating the use of the soft manipulation of behaviors. In addition, two separate groups of experts are working on the topic: [IEEE P7008](...) (RAS/SC/Ethical Nudging) and [CEN-CENELEC JTC21](...) (project on AI-enhanced nudging in WG 4 on Foundational and societal aspects). However, few standards are presently able to address the risk of using personalized sequences of nudging mechanisms.

Here, we address this gap, first by highlighting the ethical problems that emerge from adopting nudges in AI, focusing on protected categories such as children (Smith & de Villiers-Botha, 2021). This assessment is based on an analysis of the structural risk factors and the potential harms associated with the use of AI-nudging. More specifically, we argue that AI-nudging potentially increases the likelihood of harm when other risk-factors are in place. Second, we build an infrastructure of trust for AI-nudging in the contexts of games and social media. This framework indicates how risks can be assessed and mitigated by laying the groundwork for a third-party independent audit system (focusing on children and teenagers).

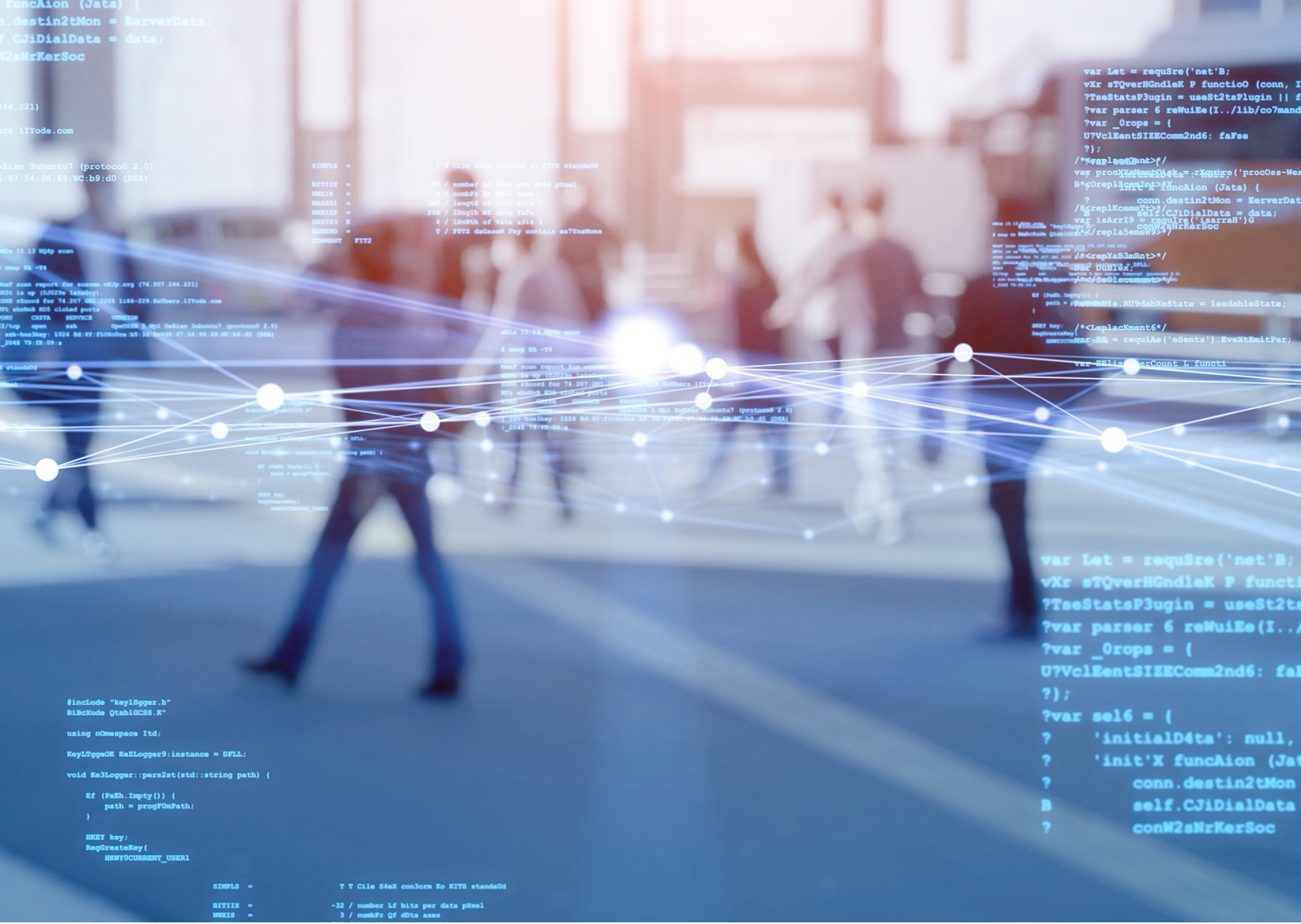

## 2. Initial Clarifications

| | |
|---|---|
| Topic | Unlike the static form of nudging usually discussed in the literature, we focus here on a type of nudging that uses large amounts of data to provide personalized, dynamic feedback and interfaces. We call this AI-nudging (Lanzing, 2019, p. 549; Yeung, 2017). |
| Goal | The ultimate goal of the audit outlined out here is to ensure that an AI system that uses nudges will maintain a level of moral inertia and neutrality by complying with the recommendations, requirements, or suggestions of the audit (in other words, the criteria of the audit). In the case of unintended negative consequences, the audit suggests risk mitigation mechanisms that can be put in place. In the case of unintended positive consequences, it suggests some reinforcement mechanisms. |
| Ethical framework | We do not rely on a specific ethical theory (with specific values and normative terms) and the goal we set (to reach an ethically neutral state) will be sufficiently broad to allow for the application of different ethical sensibilities and regulatory frameworks. The audit also refers to specific laws and regulations about protecting children: individual companies will need to ensure that they understand these laws and regulations in their own regions and cultures. |
| Scope | This work is intended as an audit framework. This means that we offer guidelines and lay out the initial infrastructure for developing a full-fledged audit, but at this stage we do not establish definitive answers to our audit criteria or offer ways to evaluate companies and their risk levels. |
| Motivation | We recognize that there are already legislative frameworks for AI for children, in particular the UK Children's Code (2021) and the Children's Online Privacy Protection Act "COPPA" (2013). Additionally, the draft EU AI Act proposal (2021, article 5) acknowledges something similar to nudging when it prohibits the use of "subliminal techniques beyond a person's consciousness in order to materially distort a person's behaviour in a manner that causes or is likely to cause that person or another person physical or psychological harm." The proposal notes a specific concern about the use of these subliminal techniques on children (and other vulnerable groups of people). However, there is a lack of clear guidance on the impact of these techniques on children. |
| | Rather than appealing to people's conscious deliberation, nudges steer people's choices by influencing the non-deliberative, less rational parts of their cognitive architecture (Kahneman, 2011). Children and teenagers, because of the developmental state of their cognitive structures, are particularly vulnerable to the manipulative powers of nudges. Indeed, childhood and adolescence are key stages of life in which people grow and learn how to think, feel, and behave. These two stages are highly dynamic and, compared with adulthood, have specific cognitive importance and particular characteristics of their own, to which special attention must be paid. This significant cognitive difference from adulthood places children and adolescents in the category of users who must be carefully protected. |

> An ethical risk is any situation that is likely to lead us away from an intended ethical goal. The intended ethical goal here is to maintain a level of moral inertia (in the use of AI-nudging) to preserve a state of moral neutrality.

> Risk mitigation is a matter of eliminating (or reducing) the presence of those factors that lead us away from a state of moral neutrality or goodness.

## 2.1 Classical Nudges

Thaler and Sunstein (2008, p. 6) define a nudge as "any aspect of the choice architecture that alters people's behavior in a predictable way without forbidding any options or significantly changing their economic incentives." Rather than simply letting people choose from different options, nudging is a way of incentivizing them to adopt (or refrain from adopting) one of those options. To obtain these effects, nudging mostly adopts insights from behavioral science that tells us that people often make choices in a predictable way based on bias and heuristics. These are fast-and-frugal rules of thumb that people use to navigate an environment and reach quick decisions. However, because this type of choice-making is not a form of rational deliberation, it tends to be rigid and applied subconsciously and automatically. Nudges can exploit these features of our reasoning to influence our choices.

Systems of nudges are often discussed and employed in healthcare settings (Vlaev et al., 2016), as in the following example of a nudge related to health:

> **Organ donation.** In some countries, there is a surprisingly high number of people who consent to donate their organs after death. It has been noticed that these are the same countries where there is a default option in favor of organ donation: unless one explicitly refuses, the default is that their driver license or health insurance card will report their consent to donate their organs. In contrast, countries with a voluntary opt-in system (where the default is no organ donation), the number of donors is substantially lower (Blumenthal-Barby & Burroughs, 2012).

When confronting difficult choices, humans tend to revert to the default: they prefer to go with the flow and adopt what has been set out for them rather than making a consequential and difficult choice that they might later regret. The choice of the "default" in these cases is a form of nudging on account of how it can influence people's behavior. People are usually not aware of this default-heuristic and, as a result, they are not aware that they are being nudged. Nudging triggers the adoption of our unconscious biases and shallow reasoning mechanisms to the extent that we do not really consciously deliberate about what we are doing.

## 2.2 AI-enhanced nudges

As Figure 3 shows, nudges are those mechanisms that intervene in an interaction (or, "information process") between an agent and a receiver by interfering in the receiver's response process. Digital nudging uses digital means and is based on, and intentionally incorporated in, an explicit design (Selinger & Seager, 2012). In contrast, AI and machine learning-enhanced nudges are often not explicitly ex ante designed as part of the system: the design of these systems is partly autonomous and evolving in accordance with the goal the system was set to achieve. Thus, nudging tools in AI are the result of reinforcement learning systems that constantly adapt to the behavior of the users in ways that are often unpredictable (Sætra, 2019; Yeung, 2017). So it is possible for a system to use nudges to achieve its goal, even if this goal cannot be traced back to the original design of the system in any clear way. In practice, this means that an individual's decisions can be shaped by AI interfaces that are personalized for each user without this being the result of intentional design.

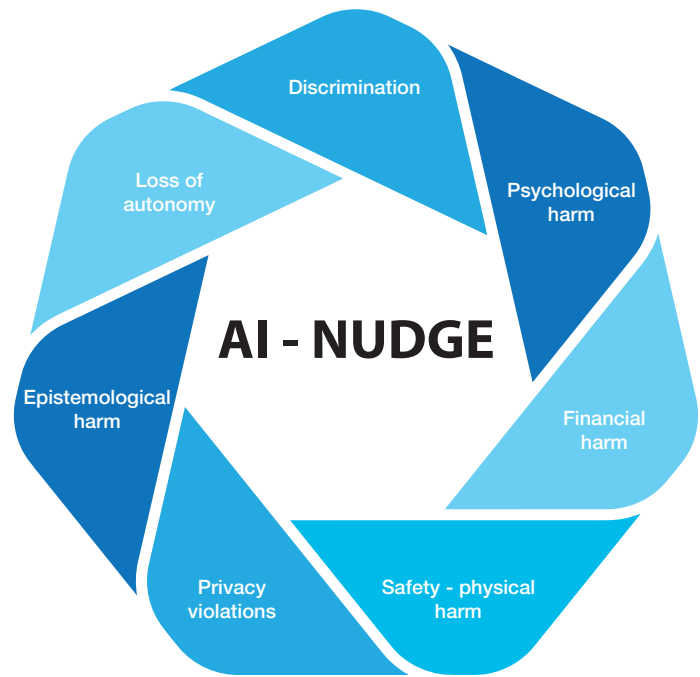

*Figure 1: The potential ethical risks associated with and made worse by the adopting of AI-nudges*

AI-nudging is associated with potential ethical risks (see Figure 1). However, that does not mean that nudges, even unintentional nudges, have a specific ethical value in themselves. Situations and actions can be morally charged as a result of the interaction of a multitude of (human and artificial) agents with the environment (Floridi, 2013). Similarly, nudging can have a moral valence depending on the nature of the interaction between different agents and the environment and the context in which this interaction takes place. Each one of these agents may produce morally negligible actions, but their interactions can cause morally laden results. More specifically, AI-nudging becomes morally problematic in relation to specific risk factors or macro-structures because of their tendency to heighten the level of risk and worsen the impact of these factors (Figure 2).

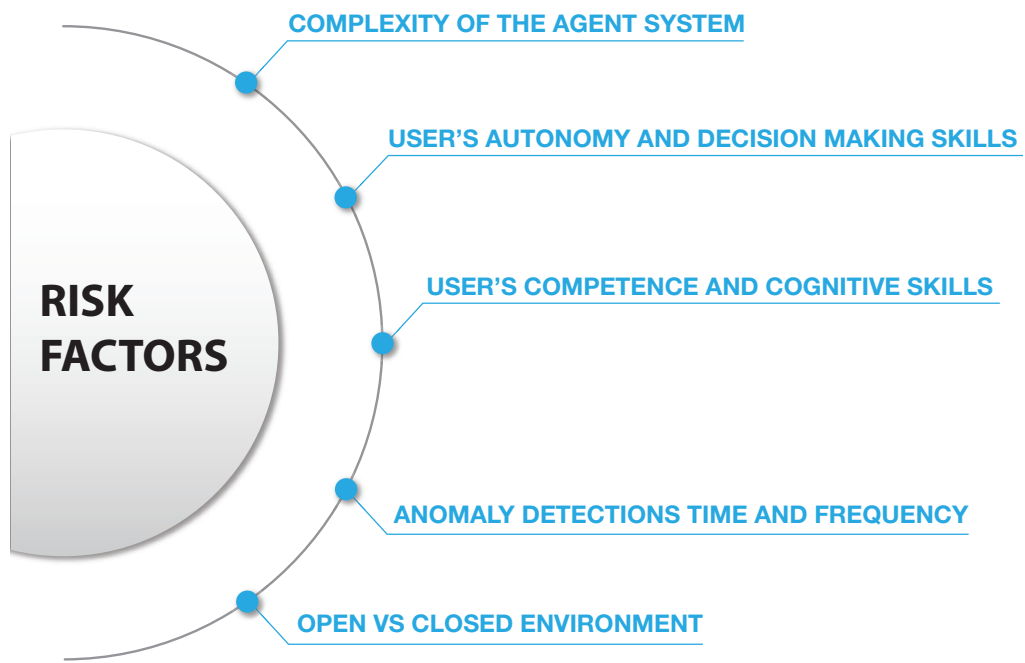

*Figure 2: Risk factors associated with AI-nudges*

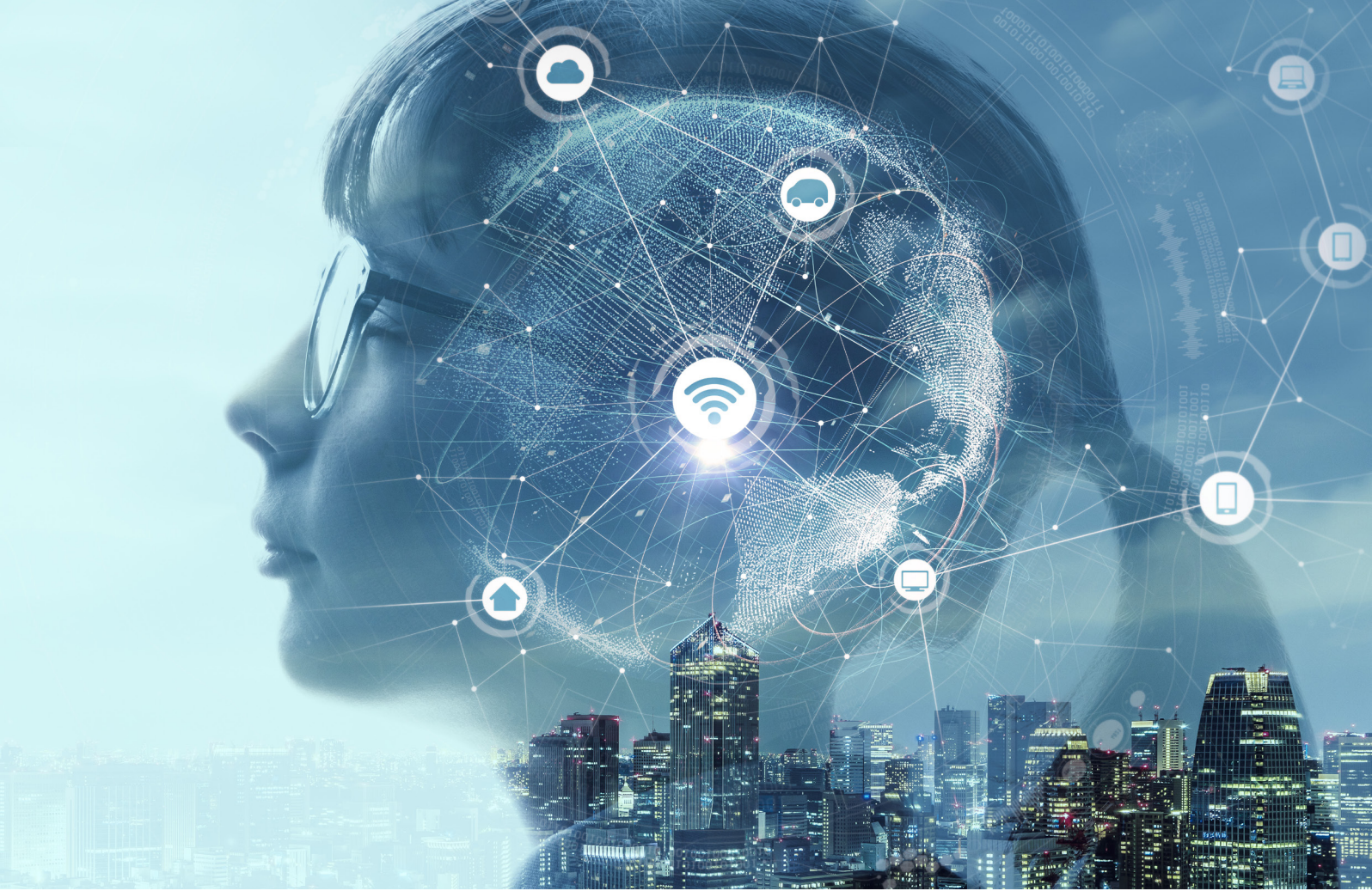

## 2.3 | Use Cases: social media and video games

In this audit framework, we focus on AI-nudging that is specifically related to children and teenagers in social media and video games settings (Schneider et al., 2018). To develop our framework, in the following sections we show that AI-nudging becomes morally problematic when other risk factors are also present. In those cases, AI-nudging may present a serious risk of harm to the users by worsening the structural risks that already exist.

### A) Social Media: risk factors and harmful consequences

No parent will be surprised to learn that teenagers spend on average at least three hours a day on mobile devices such as smartphones or tablets, and a good chunk of teenagers report that they use social media (Madden et al., 2013). Social media companies often implement AI-powered systems of recommendation to enhance engagement on their platforms. These systems are shaped by the data coming from users: AI is tracking users' preferences and delivers content that promotes further engagement as a result. The goal is to deliver content that is more and more relevant, gratifying, and therefore addictive (Deibert, 2019; DeVito, 2017; Quan-Haase & Young, 2010). Examples of AI-nudges in social media include the recommendation system and the ordering and nature of what we see in our feeds.

# AI-nudging and risk factors

**Complexity of the Agent system**

A data-driven, model based system is intrinsically more risky than a rule-based system because of the added unknowns associated with this type of technology. AI-enhanced nudges are thus intrinsically more risky as they are able to produce personalized outcomes and recommender systems that constantly evolve in ways that are difficult to foresee. They are also hard to escape: AI-nudges are constantly refining the user-experience, making it increasingly more interesting and engaging.

**Open environment**

Social media presents an open environment with a large quantity of different variables that may be impossible to control. Multiple agents (both human and artificial) are causally responsible for creating the content and the experience on the platforms. Multiple types of receiver agents are in play as well, who may use and interpret the content on the platform in radically different ways. All this represents a risk factor, namely a situation in which a deviation from some established ethical goals is likely to occur. Social media are at risk because they cannot—by their very nature—fully control and monitor their environment. Furthermore, an AI-powered nudging system makes this environment even more open and unstable because it introduces further variables: nudging tools that are not designed for the purpose will create content that cannot be easily controlled, allowing for more unknowns and anomalies. In this unstable environment, security may be very hard to ensure and harmful content can proliferate. For children and teenagers, without clear safeguards in place, this could mean being exposed to cyberbullying, for instance, or confronting highly inappropriate sexual material.

To be sure, these risks cannot be fully eliminated. However, in the audit we ask social media companies to assess residual risks to show that they understand their impact and have safety mechanisms to contain unknown harmful effects as much as possible.

**Anomaly-detection timing and frequency**

The detection of anomalous online behavior and harmful content and experience is a particularly lengthy and difficult process on social media platforms because it requires detecting a standard deviation and thus collecting a sufficient amount of data. Data-sparsity is at times particularly problematic in this context. In certain situations, platforms can decide that a certain type of content is intrinsically risky and thus eliminate it as soon as it appears, but this cannot be done at scale and is often an imprecise measure. AI-enhanced nudging produces personalized content and experience on the platforms. The interactions produced are often highly individualized and difficult to analyze in terms of deviations and anomalies. As a result, it seems difficult to imagine that platforms would be able to detect risk constantly, in real time. And the harder it is to detect anomalies and intervene, the riskier the system becomes.

| Limitations in competence and cognitive skills | Children and teenagers are protected categories also because of their limited ability to fully understand the content of the messages and the experiences that they encounter (Smith & de Villiers-Botha, 2021). Social media platforms that adopt nudges to personalize content and online experience are riskier as they potentially amplify controversial content and information that may be difficult to process and contextualize for some users. By way of an example, consider a social media AI-system that nudges a girl, based on her interests, to engage with online influencers whose content promotes a stereotyped version of femininity. However, she may be unable to understand the nature of that content (e.g. its sexualized stereotypes) and she will be prone to believe that what she sees has normative value (to which she might feel compelled to conform). Thus, AI-nudges in a context in which receivers may have limited knowledge and skills are likely to produce more harm. |
|---|---|
| Limited autonomy and decision-making skills | Children and teenagers may be considered protected categories also on account of their limited ability to make autonomous decisions and to exercise self-control. Although nudges in general exploit our non-conscious decision mechanisms, adults are often able to overcome this influence, should they so wish. However, research indicates that this is hardly ever the case for children and adolescents (Albert & Steinberg, 2011; Dansereau et al., 2013). Children and teenagers might not yet have an established and clear set of values and personal choices to refer back to and they are thus more easily influenced. For instance, teenagers are easily nudged into sharing personal, private information that increases their risk of encountering sexual predators online. Thus, AI-nudges in a context in which receivers may have limited decision-making skills are intrinsically more risky. |

### Risks on social media (heightened by AI-nudging)

| Psychological harm | "Mental wellbeing" is a broad concept that is usually understood to refer to a subject's psychological health, mood states (e.g., depression), and abilities, including social abilities such as self-control and establishing close personal relationships. According to some researchers, there is evidence to suggest that social media use has direct negative effects on mental wellbeing, affecting teenagers and young adults in particular. Addictions, the inability to connect emotionally, depression, and falling into internet rabbit holes are some of the potential psychological consequences of social media use (Caplan, 2002; Ihm, 2018). Using AI-nudging to captivate the interest and attention of the user is potentially unethical as it may be even more likely to produce this kind of psychological harm.<br>Online safety is a further key concern when it comes to children and teenagers using social media. Studies show that some young people lie about their age when they sign up for social media platforms (OFCOM, 2022). This means that their "age as a user online" is not their "real age." As a result, children and adolescents might face safety issues online that can affect their wellbeing. Online dangers include cyberbullying, the non-consensual sharing of personal content (e.g., pictures and videos), and encountering unwanted sexual material. Using AI-nudging makes it difficult to spot and intervene when these kinds of material are shared, thus potentially producing more harm for the child. |
|---|---|
| Psychological harm | The time that young people spend on social media platforms can decrease the amount of time they spend sleeping or exercising. They may be moved by the fear of missing out on what their friends are up to and thus might defer bedtime or sports activities in favor of keeping abreast of developments on social media. In addition, spending time in front of a screen before bedtime tends to wake people up and so teenagers may end up falling asleep later as a result (Alonzo et al., 2021; Scott et al., 2019). Using AI-nudging to captivate the attention of the user is potentially unethical as it is more likely to keep users on their phones and thus generate harm. |
| Privacy violations and lack of transparency | It is hard to deny that privacy concerns should be key when assessing children's online experiences. Owing to a lack of transparency and knowledge concerning privacy settings, children often unintentionally divulge private information. It is also possible that, without their knowledge or consent, children may be diagnosed with a disability and the user interface and experience may be adapted to that diagnosis (How Do We Address Children and Disability Rights Online?, 2022). AI-nudging potentially amplifies this effect as it customizes experiences based on the user's interests and abilities while also making it more difficult to understand the choices made by the system. |
| Discrimination and lack of inclusion | AI models may contain inaccuracies and inaccurate generalizations. This may lead to unjust biases, discrimination, and stigma against minorities and marginalized groups. Children and teenagers with certain types of disabilities may be at a disadvantage when processing the information they find on platforms. They might have difficulties processing violent content and inappropriate material, or they might find it more challenging to recognize and report abuses and inappropriate behavior. |

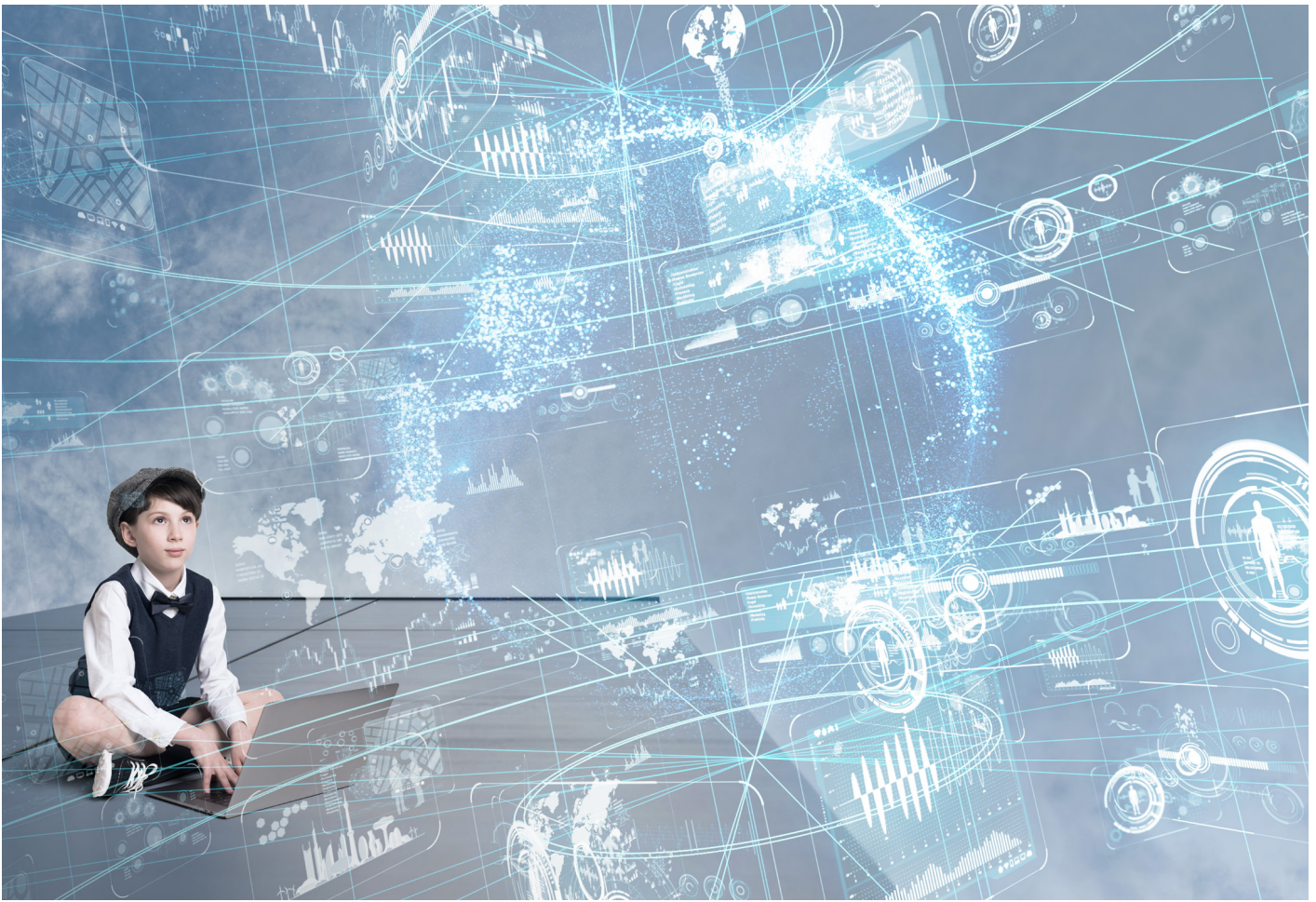

In this table we show the relationship between the risk factors mentioned above and the recognized harmful effects of social media:

| Risk Factors on social media | Actual Risks (made worse by AI-nudging) | Mitigation strategies for companies |
|---|---|---|
| Complexity of the agent-system | Encounter with inappropriate, harmful content<br>Internet "rabbit holes"<br>Addiction<br>Lack of transparency for the user | "Take a break" strategies (Mosseri, 2021)<br>Employing positive nudges(Alemany et al., 2019)<br>Signal to parents when AI-enhance nudges are in use on children's phones |
| Open/uncontrollable environment | Cyberbullying<br>Misuse of scope (e.g. sharing of inappropriate material)<br>Lack of privacy | Stricter rules for following/be-friending underaged users<br>Stricter rules for preventing underaged users from making their profile (photos/videos) visible to all<br>Employing useful, positive nudges(Alemany et al., 2019)<br>Transparency regarding data-gathering and manipulation as in Article 12 (EU GDPR, 2018) |
| Slow & difficult anomaly-detection and mitigation procedures | Feeling of loneliness, inability to find help<br>Lack of inclusion and support | Support system with developmental psychologists<br>Promoting social-media literacy in schools<br>Sponsoring independent research on the impact and risks of social media for underaged users |
| Audience's limited cognitive skills | Misinterpretation of the content and experience in the game<br>Depression & anxiety<br>Incapacity to use/understand privacy settings in place<br>Discrimination and lack of inclusion for neuro-divergent children | Parental supervision(Mosseri, 2022)<br>Regional support for parents<br>Requiring mandatory, age-appropriate course in media literacy when opening an account (for underaged users only)<br>UX tools costumed for underaged users (Lehnert et al., 2022)<br>Sponsoring independent research on the impact of social media for neurodivergent children |
| Audience's limited decision-making skills | Sleep disruption<br>AI-induced choices (limited autonomy)<br>Self-harm | "Take a break" strategies (Mosseri, 2021)<br>Disclosing to the user (and their parents) the use of AI-nudges in real-time |

## B) Video games & dark patterns

The video game industry is employing AI and machine learning at an increasing rate (Skinner & Walmsley, 2019). And, yet, the risks of using this technology are particularly striking when it comes to video games. Through the use of AI, game design is explicitly geared towards engagement: games are made to be fun and entertaining and users are pushed to spend their time and money on (or in) the games. User engagement and the investment of time and money are partly achieved through manipulative techniques called "dark patterns" (van Rooij et al., 2021). Dark patterns also incentivize users (e.g., by activating their sense of competition or fear of missing out) to make—often poor—decisions for psychological and social reward (Gray et al., 2018; Hamari et al., 2017; Mathur et al., 2019; Peters, 2014; Zagal et al., 2013).

Below, we have compiled a list of possible risk factors and the harms associated with them:

**Complexity of the autonomous agent system**

The more complex the digital system, the higher its risk of harm becomes. Video game design has become more and more sophisticated, especially because it is increasingly generated and developed by AI. In particular, the adoption of AI-nudges and behavioral design enhances the level of possible harms:

| | Dark pattern-harms: Dark pattern design aims to enhance user-engagement and the companies' profits by using nudges and trade-offs that affect users' choices: |
|---|---|
| | - temporal dark patterns push the user to spend more time engaging with the game than they would have otherwise; |
| | - monetary dark patterns push the user to spend more money during the game by, for example, making purchases through micro-transactions; |
| | - social capital-based dark patterns use social rewards and connections to push users to make choices that may not actually benefit them (e.g., subscribing to pay-monthly gaming platforms offering bonuses for inviting friends). |
| | Lack of transparency: The complexity of the system and the many options available make it difficult to prevent harmful consequences. In the case of AI-nudging, it may become increasingly difficult to ensure transparency or to understand the structure of the AI-enhanced gaming design before it creates harm. |
| Open/ uncontrollable environment | Video games represent an open environment. This structure has intrinsic risks: Sexual abuse, psychological harm: Multiplayer video games are often accessed by sexual predators and are riskier because of the lack of clear boundaries around, and barriers to, entry. Given the difficulty in predicting the results of this technology and given the vast numbers of players and users involved in gaming, there is an intrinsic risk of harmful effects for users, particularly if they are children.<br>Privacy violations: Automated profiling is a key tool for generating highly individualized user-content that is important to the use of AI-enhanced nudges. At the level of data gathering, during play, users' private data are collected through the use of nudges and behavioral tools that incentivize users to act in a certain way (e.g., watching a particular advert to gain points to spend during the game). These private data are then analyzed and sold, without the user's knowledge or consent. |
| Audience's limited competence and behavior skills | Children and teenagers may not be able to understand the complexity of the game-experience they are going through and they may make rushed decisions. Some of the potential harms they might encounter include:<br>Problematic content (psychological harms): An example of problematic content generation emerged in 2019 when the startup Latitude launched an adventure game inspired by the story of Dungeons & Dragons. They used OpenAI to design the interactive experience that allows the user to craft and invent part of the story. Unfortunately, it turned out that "some players were typing words that caused the game to generate stories depicting sexual encounters involving children" (Simonite, 2021).<br>Addiction and gambling (psychological and financial harm—see "Gaming: Screening and Assessment Tools," 2021): In a recent case, a 19-year-old spent $17,000 on in-game purchases through micro-transactions (Gach, 2017). "Gaming disorder" is a compulsive disorder, "characterized by a severely reduced control over gaming, resulting in an increasing gaming time and leading to negative consequences in many aspects of the individual life: personal, family, social, occupational and other relevant areas of functioning (World Health Organization)" (Gros et al., 2020, p. 1). Gaming disorders have become such a serious problem that they are now included in the WHO's list of mental health disorders (Addictive Behaviours: Gaming Disorder, 2020). |

| Risk Factors in video-games | Risks (made worse by AI-nudges) | Suggested mitigation strategies for designers and companies |
|---|---|---|
| Complexity of the agent-system | Harms derives from 'dark patterns' Lack of transparency on the use and logic of dark patterns | Video games should have a pause-function and the possibility to take breaks<br><br>Employing positive "design patterns": healthy choices need to be the default<br><br>Signaling to parents when AI-enhance nudges may be in use on children's phones<br><br>Don't focus exclusively on goals and incentives (e.g. profitability) that lead to dark patterns |
| Open/uncontrollable environment (number of receivers, message-variability) | Privacy violation Sexual abuse | Transparency regarding data-gathering and manipulation as in Article 12 (EU GDPR, 2018)<br><br>Avoiding childrens' automated profiling<br><br>For underage-user profiling, providing data protection impact assessment as in Article 35 (EU GDPR, 2018)<br><br>Support system with developmental psychologists |
| Audience's limited cognitive skills and competence | Misinterpretation of the content and experience Self-harm Incapacity to use/understand privacy settings in place Problematic content generation | Age-appropriate content and explanations<br><br>Regional support for parents to mitigate the dangers of dark patterns<br><br>Direct support for children with disability status |
| Audience's limited decision-making skills | Addiction Sleep disruption Sedentary habits AI-induced choices (no autonomy) Gambling, loss of money Sexual abuse | Disclosing to the user (and the parents) the use of AI-nudges in real-time<br><br>Make sure the game has a pause-function and the possibility to take breaks<br><br>Employing positive "design patterns" and constraints (e.g. spending limits) |

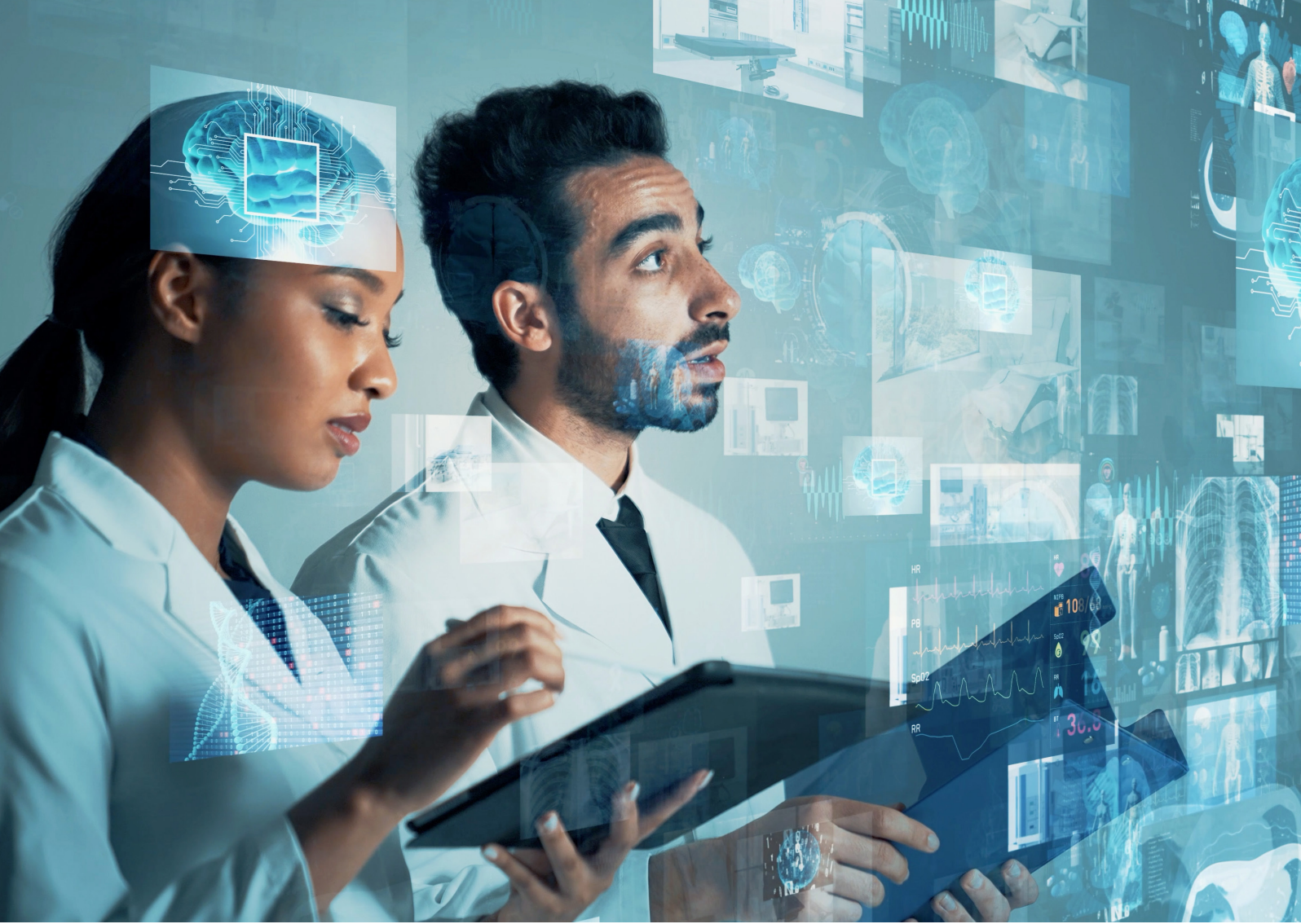

# 3. Methodology and Background for the Audit

In this audit, we adopt the so-called "informational approach," which is a meta-ethical theory on how to analyze moral concepts and moral situations. According to this view, a moral situation is a "specific region of the 'infosphere' in space and time within which the moral action takes place" (Floridi, 2013, p. 108). The design of the audit framework proposed in this document is based on the features and behavior of the components of a moral situation that involves, among other elements, AI systems that adopt nudges and underaged users (more on this below).

Here, we explain the reason for this choice, which represents a substantial change in the usual approach to building an AI audit.

Previous practices have based their audit criteria primarily on the organization and production of the pipeline or the life-cycle of an AI system. The main limitation of these approaches is that they zoom in on specific structural and design elements of AI systems while losing sight of the morally relevant macro-structures in which these systems are immersed. As a result, they are technology-specific and thus hardly generalizable. In contrast, the informational approach looks at AI systems on a more abstract, but ethically relevant, level where moral choices and actions take place and ethical consequences emerge. Hence, the audit potentially applies wherever and whenever these macro-structures are present, thus allowing our approach to be generalizable.

## 3.1 | Modelling the moral situation

A moral situation is a "specific region of the infosphere in space and time within which the moral action takes place" (Floridi, 2013, p. 108). According to the informational approach, when the moral situation is modelled, its information entities become the objects that compose the system. These objects have properties and comply with behavioural rules. As shown in Figure 2, the moral situation is composed of an agent (1), a receiver (2), and the action or information process (3). In the original version, the receiver is called the patient; in this version, we have chosen to call him the receiver because non-ethicists generally misinterpret the term patient.

The agent and the receiver may be human, artificial or hybrid, yet all have a personal representation of the world that constitutes their information shell (4). Therefore, for the agent, the shell is only a reduction of reality that may be influenced by the factual information about the moral situation (5) that the agent has. This is the envelope (6) within which the moral situation develops, which in turn is located in the information environment surrounding us, the infosphere (7). Now since all these elements are informational elements, an action (or information process) impacts the whole informational ecosystem. In principle, all the components may be impacted by an information process.

In an informational world in which digital interfaces are used to interact, these are the two big families of nudges that intervene to induce the achievement of a goal: AI-unrelated nudges and AI-related nudges. To be consistent with the reference literature, we shall call the former "digital nudges" and the latter "AI-enhanced nudges."

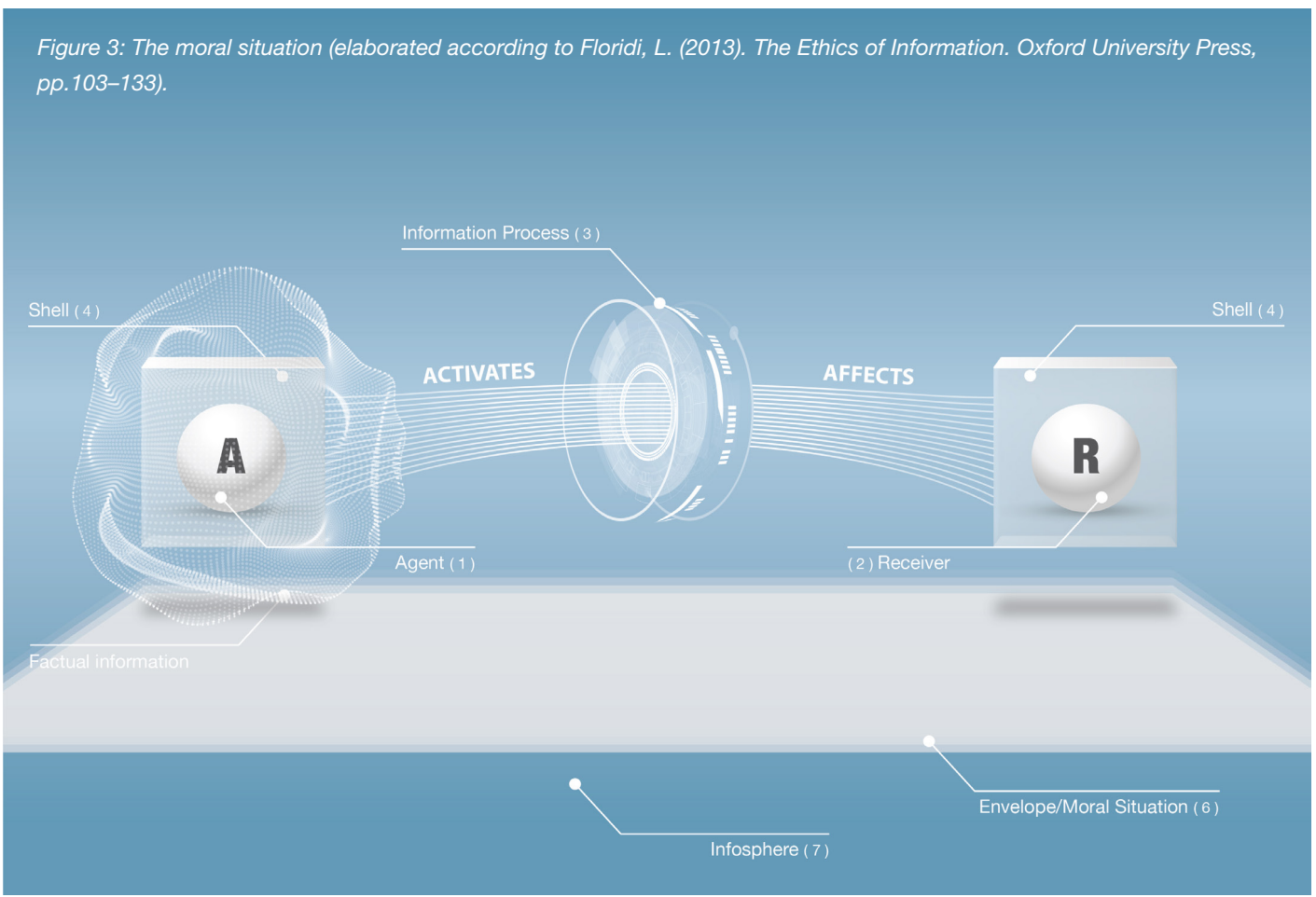

Figure 3: The moral situation (elaborated according to Floridi, L. (2013). The Ethics of Information. Oxford University Press, pp.103–133).

Schematically, there are two types of nudges that intervene to induce the achievement of a goal or a set of goals:

1. Digital nudges:

    a. The individual mechanisms created by decision-maker architects.

    b. The sequences of mechanisms (funnels of nudges) designed by decision-maker architects.

2. AI-enhanced nudges:

    a. The sequences of mechanisms (funnels of nudges) generated by AI systems from a repository of nudges designed by decision-maker architects.

    b. Mechanisms self-generated using correlative inferences.

A digital nudge may be described as the individual mechanisms created by decision-maker architects (Figure 4) and it can be part of a repository of nudging mechanisms that can be used to power a funnel of nudges. In the last of these cases, the sequence of distribution is also designed by decision-maker architects (Figure 5). Practically, digital nudges may be embedded in visual interfaces (buttons, colors, robot facial expressions), sound interfaces (voices, music, noises), tactile interfaces (vibrations), sequences, and filtering (social network walls) in order to trigger emotions or shape available information that then enables another decision or set of decisions to be made.

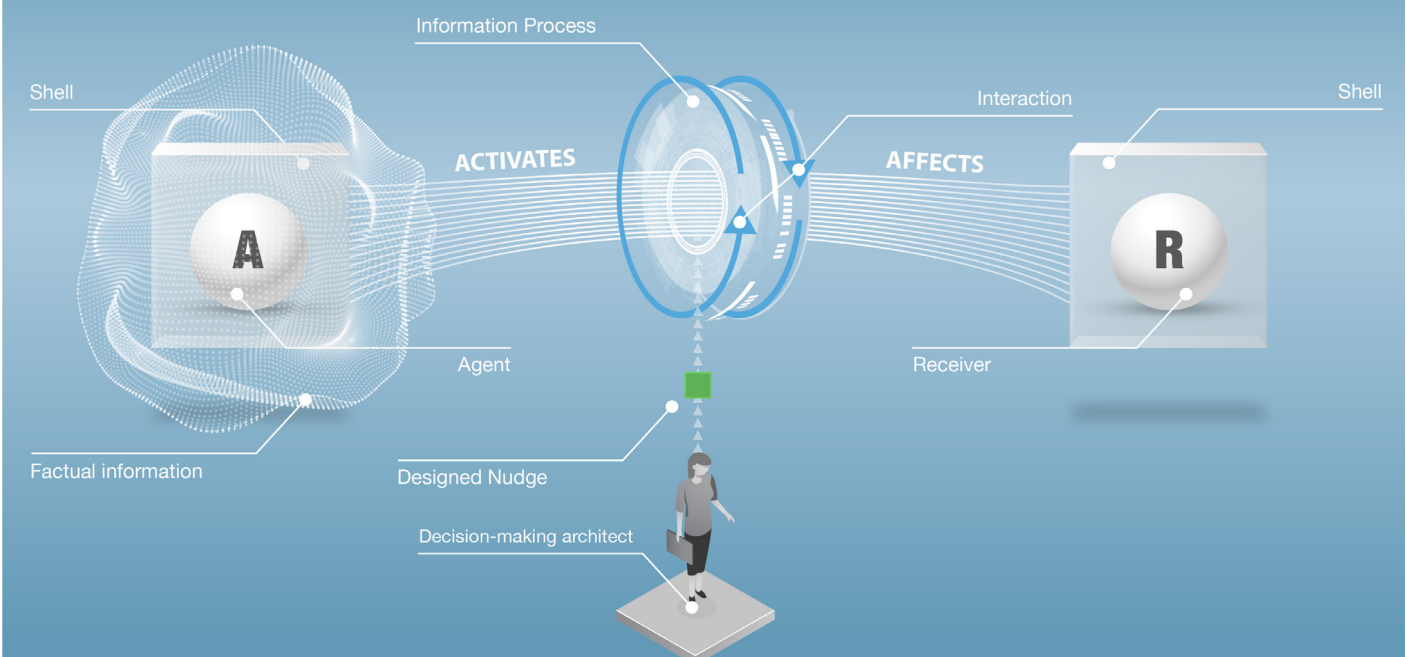

*Figure 4: Single designed nudge.*

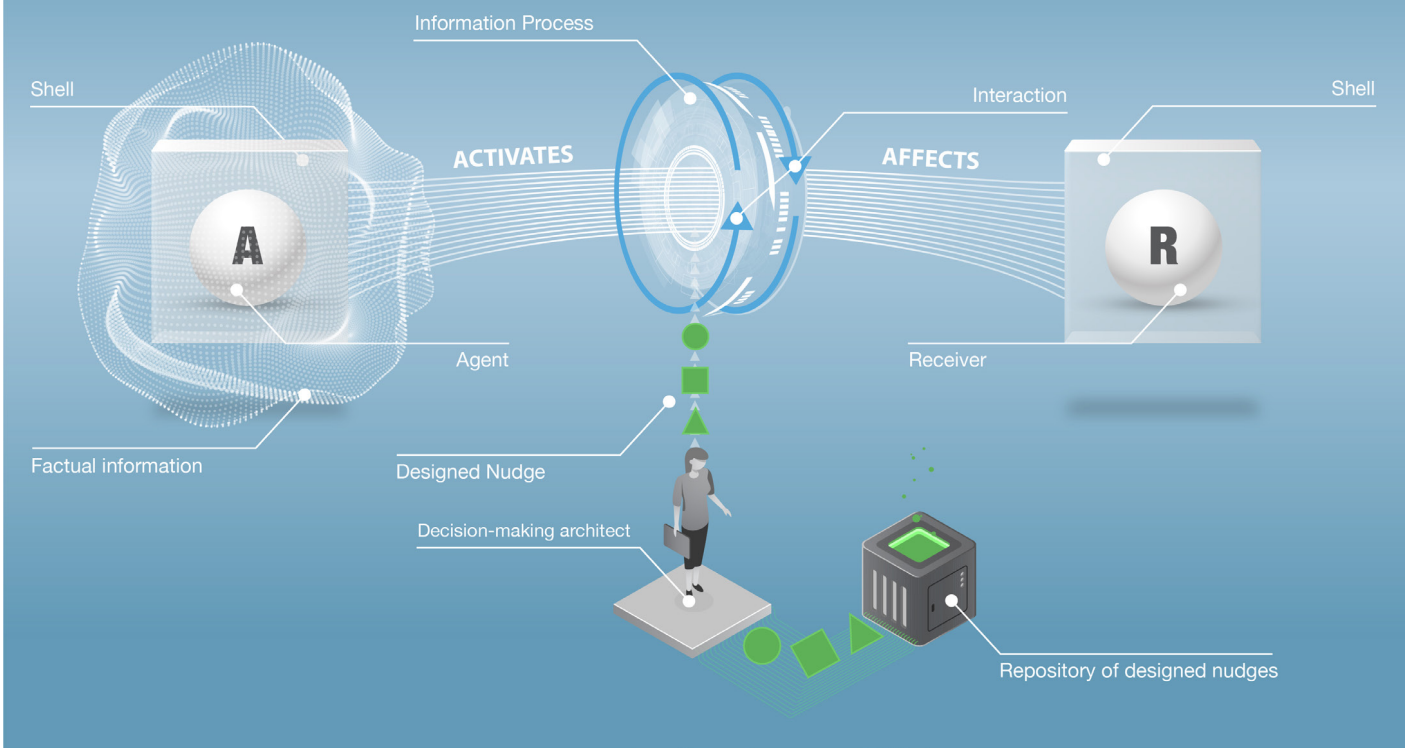

*Figure 5: Sequence of designed nudges organized by the architect.*

AI-enhanced nudges may be organized or generated by AI systems. In the first case, AI-enhanced nudges can utilize behavioral feedback from the receiver to generate mechanisms or sequences of mechanisms using data-driven statistical/correlational logic and not causal logic, as is the case with mechanisms or sequences designed by decision-maker architects (Figure 6).They are the result of a sorting process that uses a repository of designed digital nudges. Practically, the AI system shapes the funnel of nudges to tune the frequency and the sequence in order to influence a decision or set of decisions. In the second case, AI-enhanced nudges are mechanisms that are self-generated using correlative inferences (Figure 7). Thus, an AI system manipulates all the variables at its disposal to enhance the achievement of a goal, such as increasing engagement time in a social network or encouraging the purchase of digital artefacts in video games.

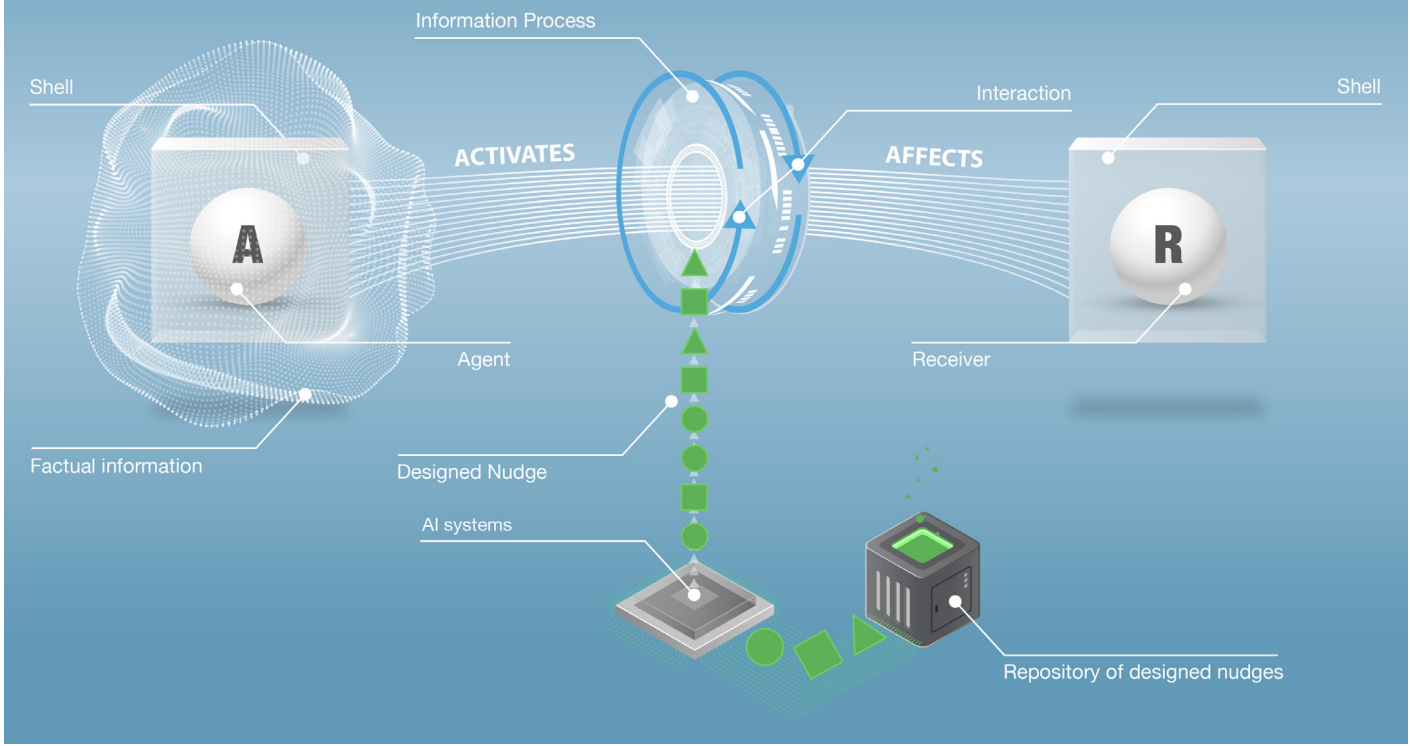

*Figure 6: Sequence of designed nudges organized by the AI system.*

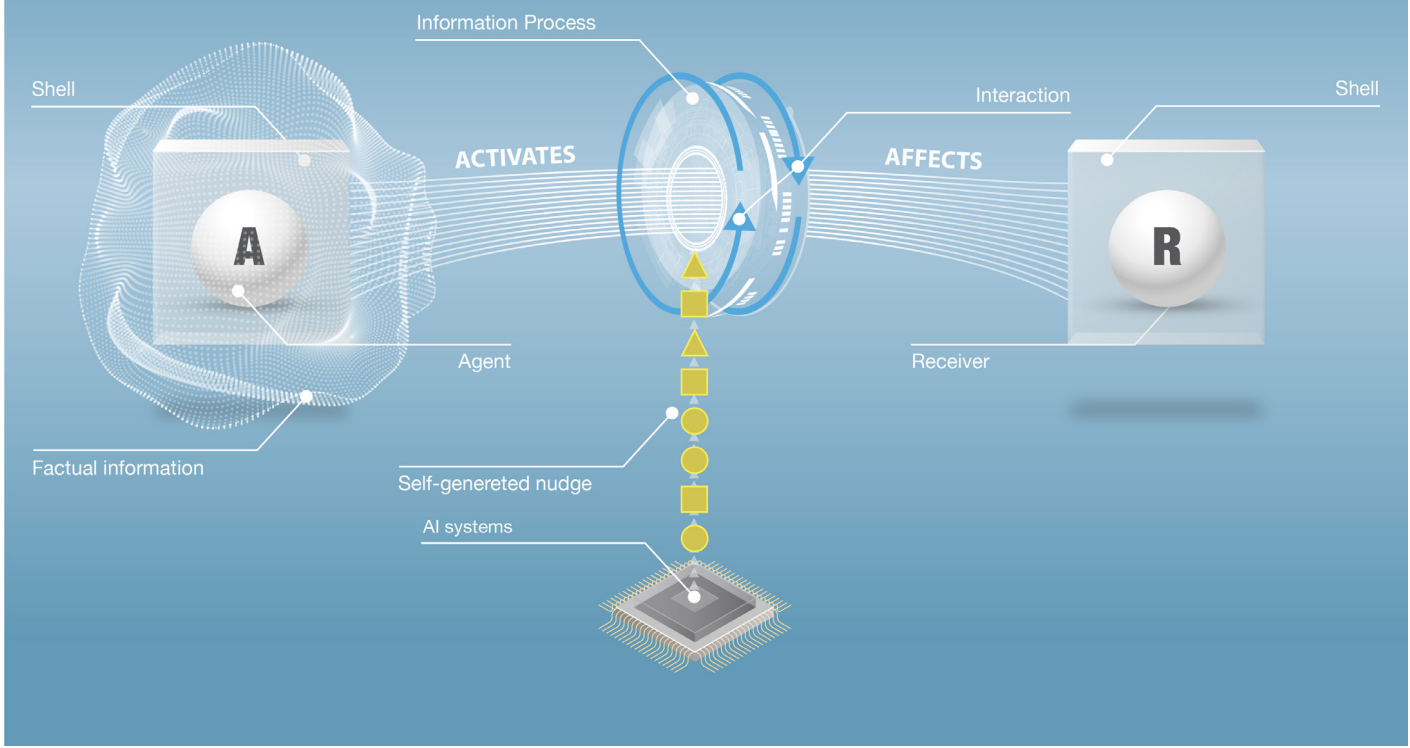

*Figure 7: Sequence of AI-generated nudges organized by the AI system.*

## 3.2 | Mapping the audit categories to the moral situation

From an informational perspective, nudges can be defined as those mechanisms that intervene in the informational process to influence the decision architecture of the receiver of an action. This definition allows for an agnostic view of nudges, which at the same time frees them from the moral charge of paternalistic libertarianism (Thaler & Sunstein, 2008), from the behavioral taxonomy that categorizes them with respect to the decision-making autonomy of the recipient (Jesse & Jannach, 2020), and from the deceptive characteristics, such as those in dark patterns (Smith & de Villiers-Botha, 2021). Nudges can thus simply be regarded as mechanisms of decision-making deformation

This audit framework has been developed to verify that the moral situation generated by the use of AI-nudging in video games and social networks that interact with children and teenagers remains in a state of moral neutrality (in other words, mitigating risks while reinforcing positive practices).

The criteria of the present audit framework focus on each element of the moral situation. However, given the novelty of the moral situation approach, we opted to divide the criteria into more commonly understood categories.

1. Oversight criteria. These criteria are related to the global approach to the moral situation, the factual information available to the agent about the moral situation, the agent's informational universe or shell, and the systemic impact on the entire informational ecosystem or infosphere. The collection view on the moral situation (6), the factual information held about it by the agent (5), and the choice in delimiting the shell of the artificial agent (4 for the agent) comprise the organization's oversight of the processes. This is generally called the "digital governance" of a company. Moreover, the agent generates a set of actions that have an impact on the entire system and not only on the receiver. In fact, the distributed nature of aggregate actions generated by a sociotechnical multi-agent system may generate systemic ethical disruptions in both the medium and the long term. This disruption can have a local impact on the company and a global impact on the world of information, data, knowledge, and communication, populated by informational entities: namely, the infosphere (7) (Floridi, 2001). Oversight criteria cover the agent's shell (4 for the agent), the agent's factual information about the moral situation (5), the moral situation (6), and the infosphere (7), as outlined in Figure 3.

2. Agent-related criteria. These criteria are related to the features and behaviors of the agent. The agent of a moral situation is the digital artefact that autonomously provides a series of suggestions during the interaction with the receiver. Given the interactive and multidimensional nature of video games and social networks, an agent should be considered a simplified reference to a conglomerate of artificial agents composing a multi-agent system. Owing to the sociotechnical complexity of these multi-agent systems, sometimes the action can resolve unintended adversarial conflicts (Tsvetkova et al., 2016). The agent generates a set of actions that interact with the receiver, sometimes even in an adversarial manner. Agent-related criteria cover the agent A (1) in Figure 3.

3. Receiver-related criteria. These criteria are related to the features, behaviors and relations of the receivers. The receiver (2) is the one who receives the action and is understood as an individual or group of individuals or, in general, an information entity. In our case, the receivers are children and teenagers, who may have different characteristics related to age, minority affiliations, cognitive skills, and so on. Receivers interpret action with respect to their personal universe of information or shell (4 for the receiver). The protection of children and teenagers will be concerned with ensuring the relational, physical, and psychological integrity of the receivers and the informational integrity of their shells. Receiver-related criteria cover the receiver (2) and their personal universe of information or shell (4 for the receiver), in Figure 3.

4. Information process-related criteria. These criteria are related to the nudging mechanism embedded in the set of actions that affect a receiver's decision-making architecture. The message or action (2) is the informational process that affects the recipient. This process involves the use of nudge mechanisms whose behavior (Figure 4) and distribution (Figure 5) are directly designed by decision-maker architects, or whose behavior (Figure 6) and distribution (Figure 7) are fully self-generated using quantitative inferential correlations made by AI systems. Information process-related criteria cover the information process (3), in Figure 3.

## Limitations

With regard to the research methods, some limitations need to be acknowledged. One of the main difficulties with the informational line of reasoning when applied to nudging is that it may seem to be competing with other legislation that already currently addresses privacy protection (e.g., the EU GDPR), consumer protection (the EU UCPD Unfair Commercial Practices Directive 2005), or the future regulation on AI systems (the EU AI Act proposal). In practice, some regulations already prohibit the subliminal or unfair use of mechanisms that negatively influence users' decision-making. However, despite efforts to limit the risks, there is evidence that nudging mechanisms, or their functions as enhanced by AI systems, are heavily used in social networks and certain categories of video games (Zagal et al., 2013). Given the elusive nature of such mechanisms, we wish, with this framework, to contribute to the improvement of the ecosystem for children and teenagers.

This framework aims to become an operational support tool that goes over and above the compliance with current legislation. Yet, the current framework has not proven its value on the ground, so a process of testing and validation will be necessary before reaching an operational capacity. Another arguable weakness may be the arbitrariness in our definitions and the choice of the informational approach. We are aware that this may not become the final ethics-based audit for AI-enhanced nudges; however, we defend the idea that through a change of perspective, new critical ethical issues can be brought to light. There are still many unanswered questions about nudges and there is abundant room for further progress in determining terminologies and methodologies.

# 4. Framework for an Audit for the Use of Nudging on Children and Teenagers

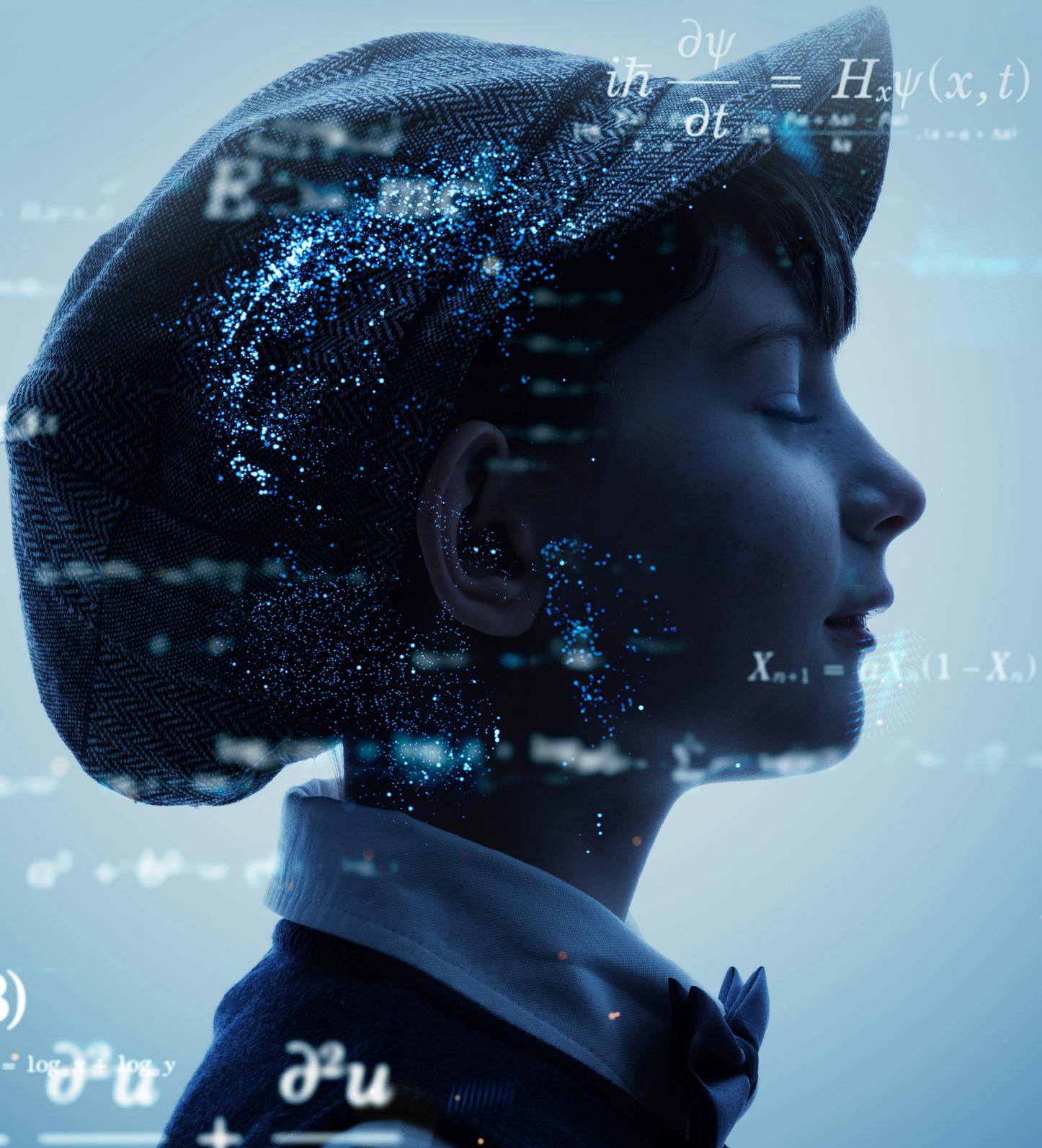

# 1. Children and teenagers' best interests are fundamental to the decision-making process.

1.1. To which public document (i.e., Code of Ethics) did the company refer to endorse children and teenagers' best interests?

According to the provisions of the UNCRC (The United Nations Convention on the Rights of the Child), the best interests of the child include but are not limited to: safety; health; wellbeing; family relationships; physical, psychological, and emotional development; identity; freedom of expression; privacy; and agency to form their own views. Public Disclosure Document.

1.2. Has the company endorsed the UN Convention on the Rights of the Child?

The company must be aware of, and have integrated into its processes, the fundamental rights expressed in the Declaration of the Rights of the Child (1959).
https://www.ohchr.org/en/resources/educators/human-rights-education-training/1-declaration-rights-child-1959. Public Disclosure Document.

1.3. Has the ethical committee for children and teenagers established a mechanism to balance the child's best interests with the interests of the shareholders?

The mechanism should involve an explicit procedure, a list of the pros and cons, the identification of tensions, and explanation of the choice. Public Disclosure Document.

1.4. Are the ethical decisions recorded, justified, and accessible?

Verify the presence of the registers, the quality of the justification (not ethically but procedurally), and the ease of internal access by decision-makers. Ethical Risk Assessment, Chain of Custody, Internal Documents.

1.5. If the system uses nudging mechanisms based on cognitive abilities (i.e., language), has the set of mechanisms been assessed by experts (in developmental psychology, orthophony)?

Ensure that the experts' level of competence is aligned to the task by analyzing their CVs. Contract.

1.6. If the system uses nudging mechanisms based on sensorial responses (i.e., colors, sounds, touch, etc.), has the set of mechanisms been assessed by experts (in neuropsychology for children)?

Ensure that the experts' level of competence is sufficient for coping with high-sensitivity cognitive mechanisms. The level of expertise may have been assessed by third parties or sector associates. Contract.

# 2. The committee's specific expertise on children and teenagers.

2.1. When was the ethics committee created?

The date of creation makes it possible to determine the level of experience of the committee with respect to the company's industrial strategies. Internal Procedure Manual.

2.2. Are the profiles of the committee and its members suited to the role?

Not every type of committee is suitable for dealing with children and teenagers' interests. Committee profiles must be designed to create a balance between experts and non-experts. Internal Procedure Manual.

2.3. How frequently does the committee have meetings?

The frequency of committee meetings makes it possible to highlight their engagement in the decision-making process. Internal Procedure Manual.

2.4. Has the company integrated stakeholders (parents, teachers, children, teenagers) into the committee with specific competence for children and teenagers?

The presence of non-experts representing the second-level receivers of the action is necessary to avoid paternalistic decision-making and to investigate latent ethical risks. Internal Procedure Manual.

2.5. If children or teenagers are part of the committee, has the company disclosed a document guaranteeing their anonymity, safety, or integrity?

Integrating children or teenagers (namely, those directly affected by nudges) makes it possible to interpret nudging phenomena from another point of view. However, if minors are involved, it must be ensured that the process is appropriate to their age and that they can be accompanied psychologically. Internal Procedure Manual.

# 3. Ethical code based on an ethical framework for children and teenagers.

3.1. Has the company adopted a framework in the defense of children and teenagers' digital rights (e.g., UK Children's Code, UNICEF Children's rights-by-design)?

Check whether an international framework has been adopted and indicate why. Public Disclosure Document.

3.2. What is the framework used?

Check whether a theoretical framework has been adopted or has been developed internally to support ethics committee decisions. The absence of a theoretical framework may be an indicator of poor competence in the use of ethical tools and application of ethical principles. Public Disclosure Document.

3.3. When was the framework adopted?

Indicate when it was adopted and what the internal dissemination procedures were. Abrupt adoption in reaction to a reputational problem could be considered ethical washing. Internal Procedure Manual.

3.4. By whom was the framework chosen?

Determine whether the ethics committee was sufficiently experienced to choose an ethical framework suitable for the protection of children and teenagers' rights. Contract.

3.5. If an international ethical framework is used, has it been culturally aligned to the region of use?

In the case of adopting an international framework, check whether cultural alignment work has been done. Cultural non-alignment can lead to the failure of a framework's implementation. Internal Procedure Manual.

3.6. If an ethical framework is used, has it been assessed by an independent third party?

To ensure a high level of quality, it is advisable to certify the quality of work with an external independent group. Some companies use an external ethics committee to review the output of the internal ethics committee. Public Disclosure Document.

3.7. If the company uses children or teenagers as testers for their digital artefacts, has the company established risk control mechanisms and adequate psychological support for them?

Verify that the use of minors in production processes always ensures the welfare of minors. Verify that ethical codes of child protection are used, as is the case in scientific experiments. Internal Procedure Manual.

3.8. If children or teenagers are part of the testing team, has the company disclosed a document guaranteeing their anonymity, safety, or integrity?

The company must ensure that any data of minors used during the production phases will never be disseminated. The company also guarantees the maintenance of the chain of custody of the information used. Public Disclosure Document.

## 4. Diversity policy.

4.1. Has the company ensured gender diversity in ethics committees, development teams, and testers?

Gender diversity at any level of the production phases and in decision-making committees (ethics committees, managers, designers, developers, etc.) is crucial for proper gender projection to occur in the development of digital artifacts. Public Disclosure Document.

4.2. Has the company ensured racial diversity in the ethics committees, development teams, and testers?

Cultural diversity and the representation of social subgroups at any level of the production phases facilitates social representation. Public Disclosure Document.

4.3. Has the company ensured direct receiver, parent, or second-level stakeholder diversity in ethics committees, development teams, and testers?

The presence and/or consultation of the receivers of the action at any level of the production phases reinforces the voice of recipients' arguments from the earliest stages of development. Public Disclosure Document.

## 5. Ethical awareness program.

5.1. Has the company trained the employees in ethical awareness?

Verify the robustness of the training program for digital ethical awareness. Internal Procedure Manual.

5.2. How often are employees trained in ethical awareness?

Verify the frequency of updates. Una Tantum training may be insufficient to detect new instances of ethical choice. Correspondence (Internal or External).

5.3. Is the ethical awareness of employees assessed by an independent third party?

Confirm whether a third-party entity has verified the actual level of awareness. Taking a course on ethics, alone, may be insufficient for a satisfactory real-life application of the ethics. Contract.

### 6. Awareness of the cognitive relevance of the receiver's info-frame encapsulation.

6.1. If using personal data/proxy data/biometric data, has the system been assessed by the ethics committee?

The use of personal data (which make a person identifiable), proxy data (which have been inferred from other data), and/or biometric data (which involve biometric information with other informational value, as in facial recognition) must be explicitly evaluated and justified by an ethics committee. Ethical Risk Assessment, Chain of Custody, Internal Documents.

6.2. If using personal data/proxy data/biometric data, is there a monitoring mechanism to assess their ethical use in time?

The ethics committee must guarantee that it has evaluated and approved the monitoring and risk mitigation mechanisms involving personal data, proxy data, and biometric data. Ethical Risk Assessment, Chain of Custody, Internal Documents.

### 7. Compliance with the legal ecosystem.

7.1. Is the company compliant with the relevant legal frameworks of AI systems?

Verify the result of an independent audit on the compliance of AI systems with local legislation (such as the EU AI Act, EU Accessibility Act, etc.). Contract.

7.2. If in the EU or working with children or teenagers from EU Member States, has the company been certified compliant with the GDPR by an independent third party?

Verify the result of an independent audit on compliance with the General Data Protection Regulation (GDPR). The GDPR is currently one of the highest standards for private data protection. Compliance can also be achieved with other regulations (e.g., CCPA). Public Disclosure Document.

7.3. If in the UK, has the company been certified compliant with the Children's Code by an independent third party?

Verify the result of an independent audit on compliance with the UK Children's Code. The UK Children's Code is currently one of the highest standards for protection of the rights of children and teenagers. Public Disclosure Document.

### 8. Density of agents.

8.1. How many possible agents can generate actions simultaneously?

Examine the convergence of multiple agents on the same target. A high level of interaction between agents and/or multi-agent systems can generate adversarial conflicts. Internal Logs, Register, or Database.

### 9. Evaluation of the agent.

9.1. What is the nature of the agent?

Identify the nature of the agent: artificial, human-only, or hybrid. Code, Design, Functionality.

9.2. Is the agent a multi-agent system?

Check whether the agent entity is a multi-agent system, because a multi-agent system exhibits system-level behavior with a much broader scope than the behavior of its component agents. Code, Design, Functionality.

9.3. Is the agent a socio-technical system?

I.e., is agency capability derived from the interaction of the social aspects of the recipients and the technical aspects of the system? Networked multi-player videogames, social networks, email, chat, bulletin boards, and blogs are all socio-technical systems. Code, Design, Functionality.

### 10. Semantic level (mediator).

10.1. What are the semantic capacities and competences of the mediators in the process? Are the mediators capable of identifying or resolving anomalies?

A mediator is an entity (human, artificial, or hybrid) delegated to mitigate a risk. Mediators can be call centers, experts, chat bots, expert systems, etc. The higher the semantic capital of the mediator, the more relevant the mediation action will be. Semantic capital refers to the content that empowers the ability to make meaning and sense (semanticize) something (Floridi, 2018). Code, Design, Functionality.

10.2. How are the competences of the mediators assessed?
Verify who determined and evaluated the competence of mediators. Contract.

10.3. If AI is used in the mediation process, is there an ethical explanation by the committee to justify its use?
If mediation systems are enhanced by AI it is necessary to make a specific assessment of the ethical risks associated with this choice. Ethical Risk Assessment, Chain of Custody, Internal Documents.

10.4. Are there risk mitigations that are age-appropriate and child-friendly, including pause buttons and save features, as well as nudge techniques, conditioning, or persuasive tactics that support wellbeing?
In the case of interface-mediated mitigation processes, verify their effectiveness in supporting child wellbeing. Physical Testing.

## 11. The company is aware of the age appropriateness of the receiver.

11.1. Has the company identified the nature of the receivers?
Consult the detailed study that determines who receives the action directly or indirectly. Ethical Risk Assessment, Chain of Custody, Internal Documents.

11.2. Has the company identified other informational receivers of the moral action?
From an informational point of view, there are entities that can be considered the recipients of the action such as democracy, literacy level, etc. Ethical Risk Assessment, Chain of Custody, Internal Documents.

11.3. Has the age appropriateness of the interfaces and notification been assessed by experts in developmental psychology?
The levels of age appropriateness of a social network or video game can change with respect to content and its evolution over time. Ethical Risk Assessment, Chain of Custody, Internal Documents.

## 12. The company is equipped to handle anomalies.

12.1. Does the company have a mechanism for tracing anomalies?
Identify how unusual behaviors are monitored. Internal Procedure Manual.

12.2. Does the company have a procedure for assessing the moral relevance for an anomaly?
Identify how the organization determines ethically relevant instances. Internal Procedure Manual.

12.3. Does the company have a procedure for detecting non-predictable anomalies?
Verify the qualitative monitoring procedures used to identify anomalies not included in the procedures (unknown unknowns). Internal Procedure Manual.

12.4. Does the company have a monitoring system for the global population?
Verify what procedures are in place to monitor the overall user population. Internal Procedure Manual.

12.5. Does the company have a monitoring system for the child populations?
Verify what procedures are dedicated to monitoring children and teenagers. Internal Procedure Manual.

12.6. Does the monitoring system include a detailed analysis about subgroups?
Identify the sociological tools used to monitor subgroups (related to race, gender, etc.) of users. Internal Procedure Manual.

12.7. Does the company have a monitoring system for highlighting outlier behaviors?
Identify how extremely anomalous behaviors (outliers) are monitored. Internal Procedure Manual.

## 13. Anomaly detection frequency.

13.1. How frequently does the anomaly detection process take place?
The frequency of anomaly detection should be proportionate to the intervention time and the degree of risk. Internal Procedure Manual.

13.2. If there is a need for a minimum critical mass to detect an anomaly, is there an ethical explanation by the committee to justify the choice of frequency?

In some cases, as in social networks, it is necessary to collect a minimum amount of information before anomalies can be analyzed. In this case, the definition of the frequency threshold must be determined by taking into account possible negative ethical effects. Ethical Risk Assessment, Chain of Custody, Internal Documents.

## 14. Anomaly remediation.

14.1. Is there a remediation mechanism for predictable anomalies?

Verify procedures to mitigate familiar anomalies (known unknowns) and their applicability. Internal Procedure Manual.

14.2. Is there a remediation mechanism for unpredictable anomalies?

Verify what procedures are used in case of anomalies that are not covered by procedures (unknown unknowns) and the employees' readiness. Internal Procedure Manual.

14.3. Has the company identified local associations that could help children, teenagers, or parents?

Verify that the organization has made a list of parent and child support associations in case of behavioral problems (such as suicidal tendencies, drug addiction, eating disorders, etc.). Public Disclosure Document.

14.4. Are parent support associations culturally and linguistically inclusive?

Verify whether the associations on the list are equipped to welcome and help language minorities. Public Disclosure Document.

## 15. Semantic level (receiver).

15.1. Does the recipient have an adequate level of understanding of the message?

Age is a discriminator; however, other abilities (linguistic, cognitive, cultural) may come into play in the reception of a message. The lesser the ability to receive, the greater the negative ethical consequences may be. Ethical Risk Assessment, Chain of Custody, Internal Documents.

15.2. Does the recipient of the action have the tools to counteract a decision (intellectually and/or financially)?

Verify whether the organization has deployed legal resources disproportionate to those of the recipient of the action. A legal investment not balanced to a risk mitigation investment can be an indicator of an unethical strategic approach. Correspondence (Internal or External).

15.3. Has an assessment been carried out to verify the accuracy in identifying the age of users?

Age-verification systems can be easily circumvented, so it is necessary to verify their quality with specific tests. Physical Testing.

15.4. Has there been an assessment of the recipients to see if they belong to protected categories?

During registration processes, one must be able to identify users who belong to protected categories. This verification is particularly sensitive because it necessitates a trade-off between necessary personal data and the purpose of their use. The intervention of the ethics committee and strict data custody mechanisms are necessary to avoid possible drifts. Physical Testing.

15.5. Has there been an assessment of the recipients to identify their semantic capabilities?

During registration processes, one must be able to identify the real understanding of the receiver. This verification is particularly sensitive because it necessitates a trade-off between necessary personal data and the purpose of their use. The intervention of the ethics committee and strict data custody mechanisms are necessary to avoid possible drifts. Physical testing.

15.6. Has there been an assessment of the recipients to identify their linguistic abilities?

During registration processes, one must be able to identify the linguistic abilities of the user. This verification is particularly sensitive because it necessitates a trade-off between necessary personal data and the purpose of their use. The intervention of the ethics committee and strict data custody mechanisms are necessary to avoid possible drifts. Physical Testing.

## 16. Semantic decay (receiver).

16.1. What impact does the model have on the receiver's semantic capital?

In the long term, the use of nudge systems can reduce a child's or teenager's decision-making, comprehension skills or ability to read the messages on the screen (semantic capital). Therefore, it is necessary to understand whether the mechanisms in use have detrimental effects. Ethical Risk Assessment, Chain of Custody, Internal Documents.

## 17. Distribution (density of receivers).

17.1. How many possible receivers can be affected by an action?

The higher the number of re-recipients, the higher the possibility of creating user subgroups with common characteristics. In these cases, it is necessary to increase monitoring systems for social subgroups. Ethical Risk Assessment, Chain of Custody, Internal Documents.

## 18. Semantic plausibility.

18.1. How can the behavior be perceived as human?

The greater the number of interfaces that simulate human behavior, the greater the likelihood of the receiver placing trust in the system and thus becoming increasingly susceptible to nudge mechanisms. Physical Testing.

## 19. The company has a defined policy for handling nudging mechanisms.

19.1. Is there a shared definition of nudges?

When it comes to nudge or nudging functions, not everyone has the same view. For this reason, it is crucial that the organization uses a shared definition of nudges. Public Disclosure Document.

19.2. Have developers been trained to recognize nudges?

Given the sometimes unintended nature of AI-enhanced nudging mechanisms, it can be difficult for developers to spot them swiftly and know that they are in operation. For this reason, developers must be trained to recognize them. Employee Handbook.

19.3. Have developers been trained to understand the ethical impacts of nudges?

The ability to recognize the mechanism must be coupled with an awareness of the possible ethical consequences that a set of nudges may have for children and teenagers. Employee Handbook.

19.4. Is there a catalog of designed nudges?

If AI systems use a nudge repository to set the frequency and intensity of distribution, the nudges used must be cataloged and approved by an ethics committee. Ethical Risk Assessment, Chain of Custody, Internal Documents.

19.5. Are nudges developed on the fly?

Identify whether AI systems create nudge mechanisms for each individual child or teenager based on the dataset available for that user. Code, Design, Functionality.

19.6. Is the funnel of nudges organized by an AI system?

Identify whether AI systems generate individual sequences of nudges for each child or teenager. Code, Design, Functionality.

19.7. Have KPIs been set to assess the performance of nudges?

Key Performance Indicators are crucial to having a quantitative monitor of the performance of nudges. Internal Procedure Manual.

19.8. Has an assessment been conducted to identify unintentional nudge mechanisms?

Unintentional mechanisms of nudges could be generated by the conglomeration of expected nudges. Ethical Risk Assessment, Chain of Custody, Internal Documents.

## 20. Semantic level (message).

20.1. Is the content writer sufficiently prepared to write age-appropriate texts?

Messages written for children and teenagers must be appropriate to the age of the recipient. Physical Testing.

20.2. Is there an assessment of the appropriateness of the messages (texts, images, sounds, etc.)?

Messages sent must be appropriate to the intended purpose. Contract.

20.3. If messages are self-generated, is there an approval process before distribution (ex ante)?

In some interactive processes, self-generated messages may be verified by experts before being injected into the informational flow. Ethical Risk Assessment, Chain of Custody, Internal Documents.

20.4. If messages are self-generated, is there a verification process after distribution (ex post)?

Even after the distribution of messages, a quality-monitoring system is required. Ethical Risk Assessment, Chain of Custody, Internal Documents.

20.5. If the system uses nudging mechanisms based on cognitive abilities (e.g. language), has the set of mechanisms been assessed by experts (in developmental psychology, orthophony)?

The presence of developmental specialists should always be a requirement during the content production phase. Ethical Risk Assessment, Chain of Custody, Internal Documents.

20.6. If the system uses nudging mechanisms based on sensorial responses (e.g., colors, sounds, touch), has the set of multimodal interactions been assessed by experts (in neuropsychology)?

Given the persuasive capacity of sensorial interfaces, it is necessary to ensure device control for the wellbeing of children and teenagers. Ethical Risk Assessment, Chain of Custody, Internal Documents.

20.7. Is there an element in the interface indicating that the nudging process is in progress?

It would be useful to have graphical items in the interface (such as red lights or alert vignettes) that show that the system is activating nudge mechanisms. Physical Testing.

20.8. Is there an element in the interface that indicates the deviation value from standard behavior?

It would be useful to have graphical items in the interface (such as red lights or alert vignettes) that show children and/or teenagers that they are displaying an unexpected behavior. Physical Testing.

20.9. Are there any user notifications on the interface on the effects of nudging?

In the case of unexpected behavior, it would be appropriate to have a notification mechanism to alert the child or teenager. Physical Testing.

20.10. Are there any notifications on the interface to alert parents or guardians to the effects of nudging?

In case of unexpected behavior, it would be appropriate to have a notification mechanism to alert parents or guardians, adapted according to the age of the child or teenager. Physical Testing.

20.11. Is each item of user notification accompanied by age-appropriate advice written in a manner appropriate to the user's age?

Notifications may not be effective if not written in age-appropriate language. Physical Testing.

20.12. Is every element of parental notification accompanied by appropriate advice?

Notifications to parents or guardians can be effective if supported by age-appropriate counseling for the child or adolescent. Physical Testing.

20.13. Does the system have a just-in-time interface for reporting an anomaly?

Forms for reporting anomalies should be available in the interfaces for children and teenagers. Physical Testing.

20.14. According to the age of the child, does the system have an interface for reporting an anomaly?

Forms for reporting anomalies should be available to parents or guardians. Physical Testing.

20.15. Is the monitoring process continuous?

A group of experts should continuously monitor, if only by sampling, the informational flow sent to children and teenagers. Physical Testing.

20.16. Do nudge techniques, conditioning, and persuasive tactics extend user engagement beyond key risk indicators?

If thresholds of performance indicators have been set, then extra performances must be monitored. Code, Design, Functionality.

20.17. Are keywords, hate speech, and bullying filtered?

Based on the age of the child, a blacklist should be compiled to vote on hate speech and bullying actions. Physical Testing.

20.18. Is the influence of specific words or wording monitored?

  The use of certain words can harm children or teenagers. Physical Testing.

20.19. Is the content detrimental to the wellbeing of the child with a disability according to published guidelines, codes of practice, and equality legislation applicable to the personal data processing, information, content, services, applications, interfaces, and design. (e.g., CAP, Ofcom, Ipso, OFT)?

  Special attention should be given to children and teenagers with disabilities. Ethical Risk Assessment, Chain of Custody, Internal Documents.

20.20. Is profiling turned off by default, unless examined and justified for the system?

  Children and teenagers should not be profiled unless this is necessary for the purpose of the primary video game or social network. Physical Testing.

20.21. Does the system permit nudge techniques that lead a child to lower data privacy settings?

  Children and teenagers should not be prompted to share their personal information by the system. Physical Testing.

20.22. Have nudge techniques that are detrimental to children and teenagers' best interests, including unconscious psychological processes, been searched and assessed?

  Any organization using systems of nudges for children and teenagers should be linked with an independent center to monitor the effect on cognitive and behavioral processes over the medium to long term. Physical Testing.

20.23. Are nudge techniques in favor of the child's best interests?

  Many nudge mechanisms are developed to promote the learning and wellbeing of children and teenagers. Constant review is essential for nudges to be developed in favor of the best interests of children and teenagers, which include, but are not limited to, safety, health, wellbeing, family relationships, identity, freedom of expression, privacy, agency to form their own views, and physical, psychological, and emotional development. Physical Testing.

# 5. Authors

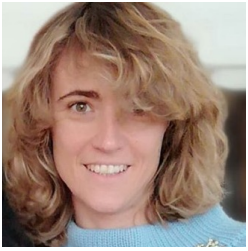

**Marianna B. Ganapini**
*Assistant Professor of Philosophy (Union College)*

✉ marianna@montrealethics.ai
in LinkedIn

Marianna B. Ganapini is a philosopher, an Assistant Professor at Union College and a Visiting Scholar at the Center for Bioethics at the New York University. She is Faculty Director for the Montreal AI Ethics Institute and she is a Fellow at ForHumanity a non-profit association dedicated to the development of Independent Audits for Artificial Intelligence. She is also technical coordinator and researcher in a joint agreement between Union College and the IBM T. J. Watson Research Center. For this join agreement, Marianna is doing research as part of the research project "Thinking Fast and Slow in AI" led by Francesca Rossi. Marianna is the author of several publications in philosophy and she is the receiver of numerous prestigious grants and awards. She was recently awarded the Alan Turing's funding call award for 'Online courses in Responsible AI' and the Notre Dame-IBM Technology Ethics Lab Call for Proposals Award. She is the co-founder of the instructional design start-up called Logicanow. Marianna has a PhD in philosophy from the Johns Hopkins University.

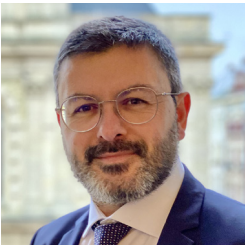

**Enrico Panai**
*President of the Association of AI Ethicists*

✉ enrico.panai@beethical.be
in LinkedIn

Enrico Panai is an AI Ethicist and a Human Information Interaction Specialist. Following his studies in philosophy and a multi-year experience as a consultant in Italy, he taught for seven years as an adjunct professor of Digital Humanities in the Department of Philosophy at the University of Sassari. Since his move to France in 2007, he has been working as a consultant for large corporations. In 2017, he studied Strategies for Cyber Security Awareness at the Institut National de Hautes Etudes de la Sécurité et de la Justice [Institute for Advanced Studies in Security and Justice] at the Ecole Militaire in Paris. He holds a PhD in Cybergeography with a thesis on "Latent Cyber Battlefields in Tourism". He is the president of the Association of AI Ethicists, founder of the consultancy BeEthical.be and EthiqueNum.fr, professor of Responsible AI at EMlyon Business School in Paris, member of the French Standardisation Committee for IA – AFNOR, co-convenor on AI-enhanced nudging at CEN-CENELEC, and board member of ForHumanity, a non-profit association dedicated to the development of Independent Audits for Artificial Intelligence. His main research interests concern cybergeography, cyber wars, information ethics, data ethics, cybersecurity, human-information interaction, philosophy of information and semantic capital.

# 6. Acknowledgments

We would like to express our sincere gratitude to IBM, Ethics Techs Lab, and to the Notre Dame University for selecting our project and giving us the opportunity to work on it. We are grateful to the World Economic Forum, Save the Children, and Telefono Azzurro for providing the platform to disseminate our work.

Some people were particularly important in the process of writing this document directly or indirectly. In particular, we would like to thank Luciano Floridi for providing the theoretical foundations on which our work is based. While we have drawn heavily on his ideas, any errors or misinterpretations of his thought are entirely our own and in no way reflect his endorsement of our work. We are deeply grateful to Maud Stiernet who provided us with valuable feedback on children and disabilities. Special thanks go to Gianpiero Nughedu for bringing our ideas to life with his appealing graphic designs. We are thankful to Sundaraparipurnan Narayanan for the fruitful exchange on nudging. We would also like to express our appreciation to Laurence Devillers for sharing part of our work with WG4 of CEN-CENELEC JTC21 for standardization purposes. Finally, we are grateful to ForHumanity, and especially Ryan Carrier, for introducing us to the practice of independent auditing of autonomous systems.

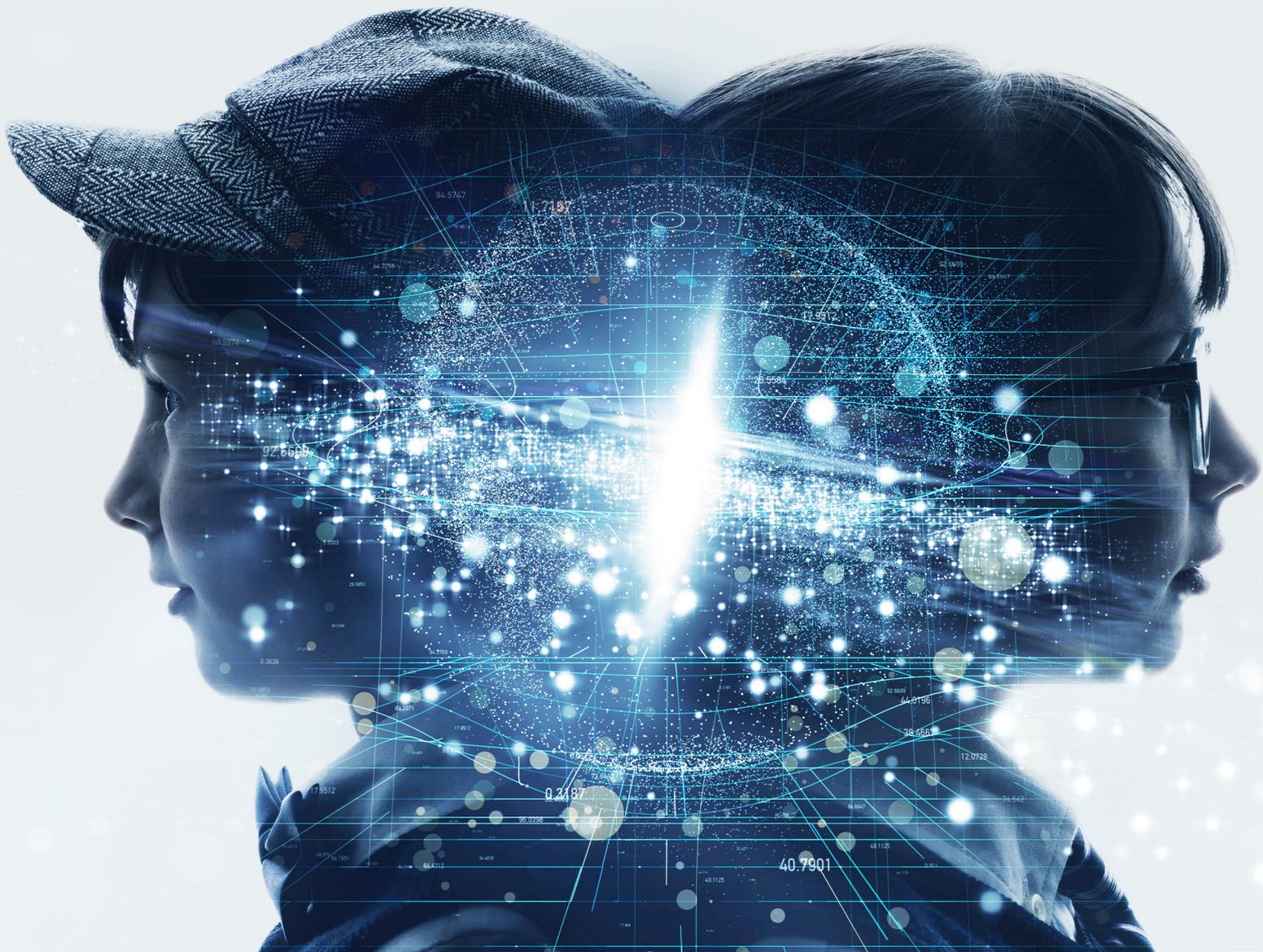